\documentstyle [12pt] {article}

\catcode`@=12
\begin{document}
\vspace{0.5in}
\oddsidemargin -.375in  
\newcount\sectionnumber 
\sectionnumber=0 
\def\be{\begin{equation}} 
\def\ee{\end{equation}}
\thispagestyle{empty}
\begin{flushright} BIHEP-TH-96-29 \\
November 1996\\
\end{flushright}
\vspace {.5in}
\begin{center}
{\Large \bf{Non-Universal Correction To $Z \rightarrow b
    \overline{b}$ \\ }}
{\Large \bf{And Single Top Quark Production at Tevatron\\}}

\vspace{.5in}
{ \rm A. Datta $^{a}$ and X. Zhang$^{b}$\\} 

{$^a$
\it Department of Physics and Astronomy, 
 Iowa State University,
Ames, Iowa
 50011\\ }

{$^b$
\it CCAST (World Laboratory), P.O. Box 8730, Beijing 100080, P.R. China \\
\it and Inst. of High Energy Phys., Academia Sinica, P.O. Box 918(4),
 Beijing 100039, P.R. China
     \\}
\vskip .5in
\end{center}  
\begin{abstract}
  New physics associated with
the heavy top quark can affect top quark production 
and the partial decay width of $Z \rightarrow b \bar b$. In this paper,
we examine the correlated effects of possible new physics on ${R_b}$ measured
at LEP I and the single
top quark production rate at Tevatron by using an effective lagrangian
technique.
We point out that certain operators in the effective lagrangian,
constrained by the measured value of $R_b$, can lead to significant and
potentially
observable
effects in single top production.
\end {abstract}
\newpage
\baselineskip 24pt

\section{\bf Introduction}
An important issue in high energy physics is to understand the mechanism of
mass generation. In the standard model, 
a fundamental complex Higgs scalar is introduced to break the electroweak
symmetry and generate masses. However,
 arguments of triviality and naturalness
suggest that the symmetry breaking sector of the
standard model is just an effective theory. The top quark,
with a mass of the order of the weak scale, is singled out to play a key role
in probing
the new physics beyond the
standard model\cite{1}.

If anomalous top quark couplings were to exist,
their effects could show up in 
 the top quark production rate \cite{2},
the partial decay width of
$Z \rightarrow b {\overline{b}}$ 
measured
at LEP \cite{1} \cite{3}, FCNC processes at low energies \cite{4}
and in
top quark decays\cite{5}.
The new experimental value for $R_b=0.2178\pm0.0011$\cite{LEP}
 is higher than the standard model
prediction of $0.2156\pm0.0005$ by $1.8\sigma$. This still
 could be a possible 
 first hint of new physics associated with
the heavy top quark \cite{Cline,Frampton}.

Recently with Chris Hill, one of the authors \cite{Hill}(X. Zhang)
  studied the correlated effects of new dynamics, which
sensitively involves the top quark, on
$R_b$ and the top pair production rate at the Tevatron. In this paper we will
study the impact of possible new physics
%, responsible for the 
% higher $R_b$ value,
 on the single
top quark production rate at the Tevatron.
% which is planed to be measured at
% $25 \%$ at Run 2 and
%$7 \%$ at Run 3\cite{9}.

It was shown in Ref\cite{SW} that the signal for single top production
in $q{\overline q}\rightarrow t {\overline b}$ 
via a virtual s-channel
$W$ is potentially observable at the Tevatron. The signal for this process
is unobservable at the LHC because of the large background from
$t{\overline t}$ production and single top
production via W-gluon fusion\cite{yuan}.
 Compared to the single top production via
W-gluon fusion the process $q{\overline q}\rightarrow t {\overline b}$
has the advantage that the cross section can be calculated reliably
because the quark and
 antiquark structure functions at the relevant values of $x$ are 
better known than the gluon structure functions that enter in the
calculation for the W-gluon cross section.  The purpose of the paper is
to show that certain types of new physics, after being constrained by the
new value of $R_b$, can still show significant effects on the single top
production rate at the Tevatron. In particular for
 operators that generate anomalous
vertices with a $q^2$ dependence one would naively expect new physics
effects in single top production to be enhanced by a factor of
$(m_t/m_Z)^2$ compared to new physics effects in $R_b$. Similar
enhancement
 effects, in models of new physics used to explain $R_b$,
could also be expected at LEP II \cite{Frampton,Kerry}.
%calculate QCD and Yukawa corrections, which are significant, to 
%the cross section 
%for $q{\overline q}\rightarrow t {\overline b}$ \cite{MW}.
% Another
%aspect that seperates $q{\overline q}\rightarrow t {\overline b}$  from
%W-gluon fusion is that the former probes the charged current interaction
%of the top quark at timelike $q^2$ while the latter does so at spacelike
%$q^2$. The processes $q{\overline q}\rightarrow t {\overline b}$ ,W-gluon
%fusion and the top decay
%($q^2\approx M_W^2$)
% are therefore complementary
%and probe the charged top quark couplings for the entire range of $q^2$.

The paper is organized as follows. In section II, we discuss the
phenomenology of $Z \rightarrow
 b \bar b$ and the single top production rate at
the Tevatron. In section III, we summarize our results.

\section{\bf Phenomenologies of $Z \rightarrow 
b \bar b$ and single top production
rate at Tevatron }

%We follow here the discussion of Ref.\cite{8}.
In Ref. \cite{Hill}, several operators in the effective Lagrangian
 relevant to $R_b$
were considered. Among them operators,
${\cal O}^{1}_{L, R}$ in the notation of Ref.\cite{Hill} are
relevant to the single top production.
 Since $b \rightarrow s \gamma$  strongly constrains
the strength of the anomalous right-handed charged current for the third
family
\cite{Fuji}, we focus here only on the operator
${\cal O}_{L}^{1}$. Explicitly \footnote{[F.1]
Operator ${\cal O}_L^1$ can be reduced to four-Fermi operator by using equation
of motion\cite{Hill}, which gives contact terms, such as ${\bar u} d {\bar t}
b$. When calculating the cross section for process $ q' {\bar q} \rightarrow
t {\bar b}$, one gets the same matrix element with or without using the
equation of motion.} 
\begin{eqnarray}
{\cal O}_{L}^{1} & =&
({\overline{\psi}}_L\gamma_{\mu}\frac{\tau^a}{2}\psi_L)
(D_{\nu}F^{\mu \nu})^{a},
\end{eqnarray}
where $F_{\mu \nu}^a$ is the SU(2) field strength, $\psi_L=(t, b)_L$,
$ D_{\mu} = \frac{1}{2}[(\stackrel{\rightarrow}{ D}_{\mu}) -
\stackrel{\leftarrow}{ D}_{\mu})]$
 and
$${\vec D_{\mu}} = { \vec \partial}_{\mu} +ig A_{\mu}^a\frac{\tau^a}{2}
+ig'B_{\mu}\frac{Y}{2}.  $$

This operator, 
 modifying 
the $Wtb$ couplings along with the $Z b{\overline b}$ vertex,
 could be 
generated in  models where the top quark has a composite
structure\cite{Georgi} and/or a soliton structure\cite{Zhang}, in the
strong ETC models\cite{ETC} and  in models where the top quark has  new strong
interactions\cite{topcolor}. It may also be generated in the
weakly interacting theories, such as SUSY and multi-Higgs models
 with relatively smaller coeffecients.

The effective lagrangian is written as\footnote{[F.2]
 We have not included the four-Fermi operators involving the top
and bottom quarks\cite{Hill},
 which can only indirectly affect $R_b$ and the
single top production rate at one-loop, but not directly as the operator,
${\cal O}_{L}^1$  does at the tree level.}
 
\begin{eqnarray}
{\cal L}^{eff} & =& 
{\cal L}^{SM} +
\frac{1}{\Lambda^2}[c_1{\cal O}_L^{1}]
,
\end{eqnarray}
where ${\cal L}^{SM}$ is the lagrangian of the
standard model and $\Lambda$ is the scale of new physics.
%These operators are expected to be generated in the models beyond the standard
%model, such as, SUSY, multi-Higgs models, ETC models and the models where 
%strong interactions are introduced and the top quark has a composite or soliton
%structure.

 The lagrangian ${\cal L}^{eff}$ generates
an effective
$Zb{\overline b}$ 
and $Wt{\overline b}$ vertices. They are
\begin{eqnarray}
{\rm{Zb{\overline b}}} & : & \frac{-ig\gamma_{\mu}}{2\cos\theta_{W}}
[g_L(1+\kappa_L)(1-\gamma_{5}) +g_R(1+\gamma_5)] , \nonumber\\
g_L &=&-\frac{1}{2} + \frac{1}{3}\sin^2\theta_{W} , \nonumber\\
g_R &=& \frac{1}{3}\sin^2\theta_{W} , \nonumber\\
\kappa_L &= & \frac{c_1M_Z^2\cos^2\theta_{W}}{2gg_L\Lambda^2} ; \
\end{eqnarray}                                   
and
 \begin{eqnarray}
{\rm{Wt{\overline b}}} & : &V_{tb} \frac{-ig}{2\sqrt{2}}
[F_1\gamma_{\mu}(1-\gamma_{5}) ]
, \nonumber\\
F_1 &=& 1-\frac{c_1q^2 }{g\Lambda^2}, \nonumber\\
\end{eqnarray}
where
$q=p_t+p_{\bar b}$ is the momentum of the W.

We now use the experimental value of $R_b$
to constrain the parameters associated with the higher dimension
operator, and
 then
calculate its correction to the single top production rate.
 In the effective lagrangian  $c_1/\Lambda^2$
can be extracted by using the formula\cite{Hill}
\begin{eqnarray}
\kappa_L & =& \frac{(R_b-R_b^0)}{R_b^0}\frac{g_L^2 +g_R^2}{2g_L^2}\
\end{eqnarray}
where $R_b$ and $R_b^0$ are the experimental value
and the standard model prediction respectively.
%and so is a free parameter. 
The cross section for $p {\overline p}\rightarrow t{\overline b} X$
is given by
\begin{eqnarray}
\sigma(p {\overline p}\rightarrow t{\overline b} X) & = &
\int dx_1dx_2[u(x_1){\overline d}(x_2)+ u(x_2){\overline d}(x_1)]
\sigma(u{\overline d}\rightarrow t{\overline b}) . \
\end{eqnarray}
 Here $u(x_i)$, ${\overline d}(x_i)$ are the $ u $ and the ${\overline
d}$ structure functions,
 $x_1$ and $x_2$ are the parton momentum fractions and the indices
 $i=1$ and $i=2$ refer to the proton and the antiproton.
For the process
$$u(p_1) + {\overline d}(p_2) \rightarrow W^* \rightarrow {\overline
b}(p_3) + t(p_4) , $$
the spin and color averaged matrix
element squared  at the partonic level
is given by ( with $V_{tb}=V_{ud}\approx 1$)
\begin{eqnarray}
|M|^2 & = & 32 G_F^2 \frac{M_W^4}{(q^2-M_W^2)^2}
[F_1^2 (p_1\cdot p_3)(p_2\cdot p_4) ],
\end{eqnarray}
 We use the MRSA' structure functions, given in
Ref.\cite{structure}, for our numerical calculation.
 In Fig. 1, we plot $\Delta \sigma/\sigma$ vs $R_b$ where $\Delta \sigma$
 is the change in the
single top production cross section in the presence of higher
dimensional operators in the lagrangian and $\sigma$ is the standard
model cross section for single top production.

 One can see from the figure
that if one requires the new physics to bring the prediction of $R_b$ to
 the central value of the experimental data, $R_b=0.2178$, then its correction
to the single top rate is around $13 \%$. Furthermore, if we 
fit $R_b$ to the experimental value within
1 $\sigma$ then 
the correction due to new physics
to the single top production rate
could be $6 \sim 20 \%$ 
 \footnote{[F.3] We have
 not included the QCD and Yukawa corrections\cite{MW} to the single top quark
production rate. They will enhance the total rate, but not change
 the
percentage of the new physics correction to the cross section.
}.
\section{\bf Conclusion}

In this paper, we have studied the correlated effects of new physics on
$R_b$ and the cross section of the single top production at the Tevatron. 
Our results show
 that the correction to the single top quark production rate due to the
new physics responsible for $R_b$ could be  $6-20 \%$ at the
 Tevatron.
Given that the single top cross section (via $q^\prime {\bar q} \rightarrow t
{\bar b}$ ) can be measured at Tevatron Run 2 and Run 3 with a precision of
 $27 \%$ and
$8 \%$ respectively\cite{Ann}, our study here provides a theoretical argument
for such a measurement in order to probe new physics beyond the standard
model.
                              
{\bf Acknowledgment:} We would like to thank Fermilab
 for hospitality, where this work was initiated. We acknowledge useful
discussions with
Chris Hill.
 This work was supported in part
by DOE contract number DE-FG02-92ER40730(A. Datta)  
 and by Chinese National Science Foundation (X. Zhang).

\newpage                                

\section{Figure Caption}
\begin{itemize}
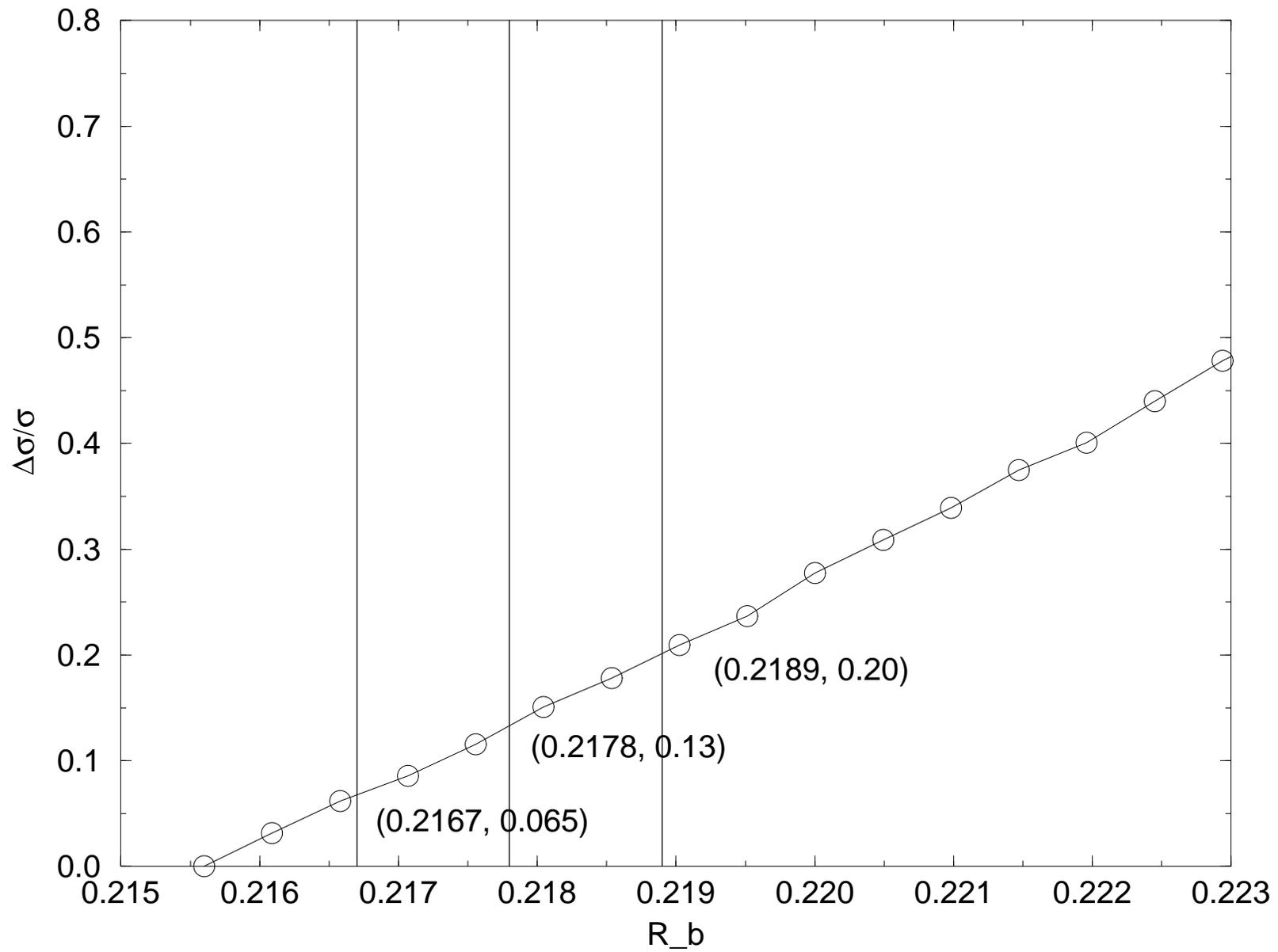
\item [{\bf Fig.1 :}] The plot shows  $\Delta \sigma/\sigma$ vs $R_b$ 
where $\Delta \sigma$
 is the change in the
single top quark production cross section in the presence of higher
dimensional operators and $\sigma$ is the standard
model prediction. The numbers shown in the plot represent
$(R_b,\Delta \sigma / \sigma)$.
\end{itemize}
\end{document}